\documentclass[showpacs,twocolumn,preprintnumbers,amsmath,amssymb,prl,epsf,superscriptaddress]{revtex4-1}
\usepackage{graphicx}
\usepackage{nicefrac}
\usepackage{color}
\usepackage[squaren]{SIunits}

\begin{document}
\title{Polarization-Entangled Light Pulses of $10^5$ Photons}
\author{Timur Sh.~Iskhakov}
\affiliation{Max Planck Institute for the Science of Light,
G\"unther-Scharowsky-Stra\ss{}e 1/Bau 24, 91058 Erlangen, Germany}
\author{Ivan~N.~Agafonov}
\affiliation{Department of Physics, M.V.Lomonosov Moscow State University, \\ Leninskie Gory, 119991 Moscow,
Russia}
\author{Maria~V.~Chekhova}
\affiliation{Max Planck Institute for the Science of Light, G\"unther-Scharowsky-Stra\ss{}e 1/Bau 24, 91058
Erlangen, Germany} \affiliation{Department of Physics, M.V.Lomonosov Moscow State University, \\ Leninskie Gory,
119991 Moscow, Russia}
\author{Gerd Leuchs}
\affiliation{Max Planck Institute for the Science of Light, G\"unther-Scharowsky-Stra\ss{}e 1/Bau 24, 91058
Erlangen, Germany} \affiliation{University of Erlangen-N\"urnberg, Staudtstrasse 7/B2, 91058 Erlangen, Germany}
\vspace{-10mm}
\begin{abstract}
 We experimentally demonstrate polarization entanglement for squeezed vacuum pulses containing more than $10^5$ photons. We also study photon-number entanglement by calculating the Schmidt number and measuring its operational counterpart. Theoretically, our pulses are the more entangled the brighter they are. This promises  important applications in quantum technologies, especially photonic quantum gates and quantum memories.
\end{abstract}
\pacs{03.65.Ud, 03.67.-a, 42.50.Lc}
\maketitle

Entanglement is the signature of the quantum world. One part of an entangled system has its properties fully undefined yet fully correlated with the properties of its counterpart~\cite{DNK98}. Can this behavior be observed for large objects? Recently, entanglement was discovered for macroscopic material systems~\cite{BEC,Polzik}. It is tempting to observe it for bright photonic states~\cite{macroqubits,Gisin} because bright light is much more efficient in interactions than single photons. For bright squeezed vacuum, very different from usual squeezed light, entanglement was discussed theoretically~\cite{Drummond,Reid,Gatti,Durkin,Simon} but never tested experimentally. Coincidence measurements could only reveal entanglement for up to 12 photons~\cite{Eisenberg}.

The clue to the observation of entanglement for bright squeezed vacuum is in registering, instead of single photons and coincidences~\cite{Durkin,Eisenberg,Lamas-Linares}, fluctuations of macroscopic intensities and measuring the variances of intensity differences~\cite{Macrobell}. A great advantage of this measurement is that it is robust against the multimode detection. In our experiment, by applying this technique to entangled bright squeezed vacuum~\cite{Braunstein} we test a multimode separability condition formulated in terms of variances.

The states of entangled squeezed vacuum, also known as macroscopic Bell states, can be obtained via parametric down-conversion (PDC) in two type-I nonlinear crystals~\cite{Macrobell}. For example, the singlet state is generated by the Hamiltonian~\cite{Simon,Eisenberg,Gatti,Macrobell}
\begin{equation}
\hat{H}=i\hbar G(a_{H}^{\dagger}b_{V}^{\dagger}-
a_{V}^{\dagger}b_{H}^{\dagger})+\hbox{h.c.},
 \label{Ham}
\end{equation}
where $a^{\dagger},b^{\dagger}$ are photon creation operators in beams $A,B$, which in our case have the same direction but different wavelengths
$\lambda_A,\lambda_B$. The subscripts $H,V$ stand for the horizontal and vertical polarization, and the parameter $G$ depends on the crystal properties and the pump power. The state can be written as a Fock-state expansion~\cite{Simon},
\begin{equation}
|\Psi^{(-)}_{mac}\rangle=\frac{1}{\cosh^2\Gamma}\sum_{N=0}^{\infty}\sqrt{N+1}\tanh^N\Gamma|\Psi_-^{(N)}\rangle,
 \label{state}
\end{equation}
where
$|\Psi_-^{(N)}\rangle\equiv\frac{1}{\sqrt{N+1}}\frac{1}{N!}(a_{H}^{\dagger}b_{V}^{\dagger}-
a_{V}^{\dagger}b_{H}^{\dagger})^N|0\rangle$ and $\Gamma$ is the parametric gain.

\begin{figure}[h]
\begin{center}
\includegraphics[width=0.3\textwidth]{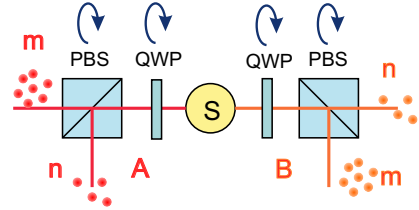}
\caption{Quantum correlations of the macroscopic singlet Bell state. The source S emits light pulses into beams A and B (spatially separated for clarity). In each pulse, photon numbers emitted into any orthogonal polarization modes in the two beams are random but exactly equal.}\label{cor}
\end{center}
\end{figure}

 Clearly, photon numbers in beams A and B are exactly the same. Moreover, if polarization beamsplitters are placed into the beams, the number of transmitted photons in beam A will be exactly equal to the number of reflected photons in beam B, and vice versa (Fig.~\ref{cor}).
Correlations will be maintained at any orientations of the polarizers, or quarter-wave plates inserted in front of them, as long as they are the same for both beams. This is because the operator expression in $|\Psi_-^{(N)}\rangle$ is invariant to any polarization transformation. Note that such perfect correlations are only manifested by squeezed vacuum and not by displaced squeezed states, which contain a huge coherent component and only a small proportion of squeezed vacuum.

Such polarization correlations are similar to the ones manifested by two-photon Bell states but involve far larger photon numbers. Besides their fundamental interest, they are important for quantum information protocols based on light-light (quantum gates) and light-matter (quantum memory) interactions. One can mention quantum memory based on bright squeezed vacuum~\cite{qmemory} and up-conversion of such states~\cite{diTrapani}. The latter suggests a new field of research, nonlinear optics of entangled states.

\textit{\textbf{Separability condition.}} How can we prove in experiment that the state (\ref{state}) is entangled? According to  Ref.~\cite{Simon}, if a bipartite system containing two macroscopic light beams $A,B$ (Fig.~\ref{cor}) is separable, it satisfies a certain condition. Violation of this condition indicates that the state is non-separable (entangled if it is pure).

To simplify comparison with experiment, we derive a necessary condition of separability in terms of the Stokes parameters and their variances~\cite{supplementary}. This approach enables us to prove a stronger condition than the one of Ref.~\cite{Simon}. It is important that our consideration is also valid for multimode beams.

As shown in Section A of the supplementary material, for a separable state, the sum of the three Stokes variances $\Delta S_i^2$, $i=1,2,3$, cannot be smaller than twice the total photon number $\langle\hat{S}_0\rangle$,
\begin{equation}
\sum_{i=1}^3\Delta S_i^2/\langle\hat{S}_0\rangle\ge2. \label{separability1}
\end{equation}
A similar inequality was proved for atomic ensembles in Ref.~\cite{Toth}.

Inequality (\ref{separability1}) is often mentioned as one of the uncertainty relations in polarization quantum optics (see, for instance, \cite{Klyshko,Klimov}). Indeed, it follows directly from the well-known equality for the Stokes operators $\hat{S}_i$~\cite{Klyshko}, $\hat{S}_1^2+\hat{S}_2^2+\hat{S}_3^2=\hat{S}_{0}(\hat{S}_{0}+2)$.
 It should be noted, however, that this operator equality holds true only in the case where, apart of the two polarization modes, the light beam contains only a single frequency and angular mode~\cite{Karas,Karas2}.  Thus, inequality (\ref{separability1}) is not of general validity. In fact, it is a \emph{necessary condition of separability}. Its violation indicates that a beam is non-separable, i.e., is \emph{a sufficient condition of non-separability}. As we show below, Eq.~(\ref{separability1}) is violated in our experiment.

\begin{figure}[h]
\includegraphics[width=0.5\textwidth]{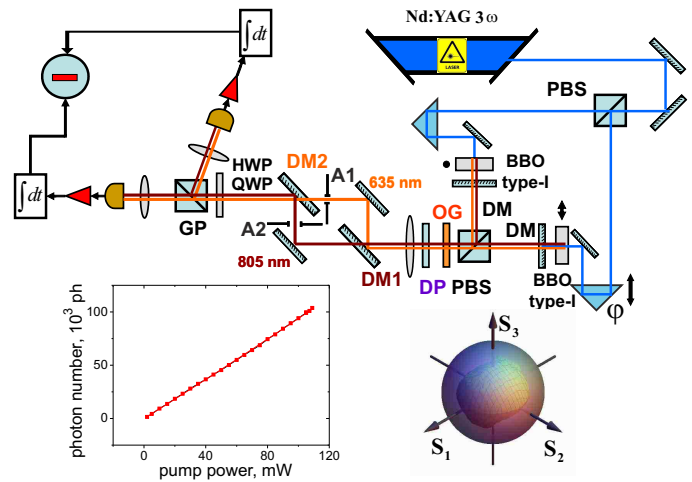} \caption{ Top: the experimental setup. Orthogonally polarized squeezed vacuums are generated in two type-I BBO crystals and overlapped at a polarizing beamsplitter (PBS); the residual pump is eliminated by dichroic mirrors (DM) and a long-pass filter OG. The interferometer is balanced using a trombone prism. A dichroic plate
(DP) is inserted for producing the macroscopic singlet state. Two apertures (A1,A2) placed in the focal plane of a lens select the angular spectra of the beams at two wavelengths, separated by a dichroic mirror DM1 and joined together by another dichroic mirror DM2. The measurement part also includes a Glan prism
(GP), a half-wave plate (HWP) or a quarter-wave plate (QWP), and two detectors. Bottom left: number of photons per pulse versus the pump power. Bottom right: variance of the Stokes observable  plotted versus the direction in 3D (the object inside the sphere). The outer sphere shows the shot-noise level.} \label{setup}
\end{figure}

\textit{\textbf{The experiment}} was performed with the macroscopic singlet Bell state $|\Psi^{(-)}_{mac}\rangle$, similar to the one considered in~\cite{Simon,Gatti,Eisenberg,Durkin}. The setup (Fig.~\ref{setup}) is described in detail in Refs.~\cite{Macrobell,supplementary}.

Theoretically, the singlet state $|\Psi^{(-)}_{mac}\rangle$ has three Stokes parameters equal to zero, $\langle \hat{S}_{1,2,3}\rangle=0$, as well as the corresponding variances, $\Delta S_{1,2,3}^2=0$, and higher-order moments~\cite{Karas}. Thus, in theory condition (\ref{separability1}) is always violated, as its left-hand side is zero. In practice, achieving a zero variance of any Stokes observable is impossible. The noise is caused by the inevitable losses (including the non-ideal quantum efficiency of the detectors) and the imperfect mode matching. To optimize the mode matching, beams $A,B$ are filtered in the angle separately.

Testing condition (\ref{separability1}) requires the measurement of variances for three Stokes observables and the total photon number $\langle \hat{S}_0\rangle$, which is the shot-noise level.
The variances of $S_{1,2,3}$ and the mean photon number $\langle \hat{S}_0\rangle$ were measured by analyzing the statistics over $20000$ pulses~\cite{Iskhakov}.
Typical photon numbers per pulse were $10^5$ (Fig.~\ref{setup}). This was due to a very large number of modes collected. It is known that collinear type-I phase matching is characterized by a large number of angular Schmidt modes~\cite{Fedorov}. By accepting, with our angular apertures, nearly whole angular spectra at wavelengths $\lambda_A,\lambda_B$, we collected about $10^4$ angular modes and $10^2$ frequency modes~\cite{two-color}. At the same time, the number of photons per mode was mesoscopic. The bottom-left part of Fig.~\ref{setup} shows the output/input characteristic of the down-converter; the dependence is almost linear and its fit yields the maximum gain $0.33\pm 0.06$ corresponding to the number of photons per mode $0.12\pm0.04$. However, condition~(\ref{separability1}) is invariant to the number of modes~\cite{supplementary}. This is why it is suitable for testing multimode states; on the contrary, traditional Wigner-function measurement requires single-mode states and is therefore not applicable here. Besides, measurement of the photon-number variance for squeezed vacuum has been shown to be invariant to the gain, at least up to values $\Gamma\sim 2$~\cite{two-color}.

\begin{figure}[h]
\begin{center}
\includegraphics[width=0.4\textwidth]{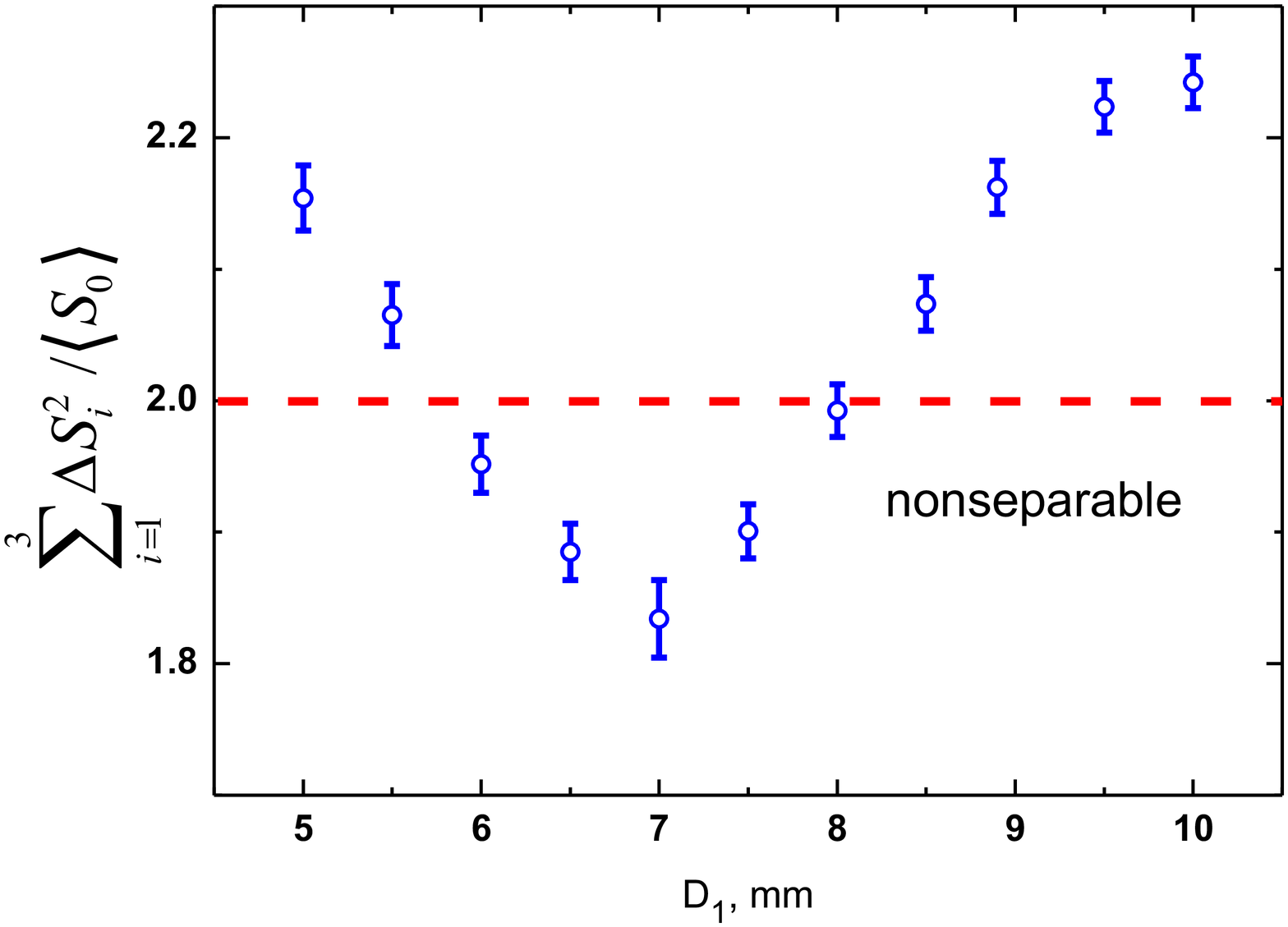}
\caption{The left-hand part of the separability condition~(\ref{separability1}) versus the diameter of aperture A1, the other aperture diameter being $8.9$ mm. As expected, the best noise suppression is observed for $D_1$ satisfying the condition $D_1/D_2=\lambda_A/\lambda_B$~\cite{two-color}, for which the mode matching is optimal. The dashed line is the boundary set by the separability condition.}\label{sum_NRF}
\end{center}
\end{figure}

Figure~\ref{sum_NRF} shows the left-hand side of inequality (\ref{separability1}) plotted against the diameter of the A1 aperture, $D_1$. For all points below the dashed line, the necessary condition of separability is violated. We see that with the transverse modes properly matched, it is violated by more than $5$ standard deviations.

Hence, we have experimentally demonstrated that the macroscopic singlet Bell state is polarization non-separable. Note that it is prepared pure, and only imperfections of the detection setup make it mix with the vacuum. A pure non-separable state should be able to violate Bell's inequalities~\cite{Gisin2} but apparently a new form of Bell's inequality should be derived for this state.

\textit{\textbf{Photon-number entanglement}} of the singlet Bell state (\ref{state}) can be characterized by noticing that the state can be rewritten as a product of two Schmidt decompositions in the Fock basis,
\begin{eqnarray}
|\Psi^{(-)}_{mac}\rangle=|\Psi_1\rangle\bigotimes|\Psi_2\rangle,\nonumber\\
|\Psi_1\rangle\equiv\sum_{n=0}^{\infty}\sqrt{\lambda_n}|n\rangle_{AH}|n\rangle_{BV},\nonumber\\
|\Psi_2\rangle\equiv\sum_{m=0}^{\infty}(-1)^m\sqrt{\lambda_m}|m\rangle_{AV}|m\rangle_{BH}.
 \nonumber
\end{eqnarray}
Here, $\lambda_n\equiv\tanh^{2n}\Gamma/\cosh^2\Gamma$ and the notation
$|n\rangle_{AH}$ means a Fock state in beam $A$ with $n$ photons
in the horizontally polarized mode. The notation for beam $B$ and mode $V$ is similar. Clearly, the state can be represented as a product of two entangled states, one of them involving modes $AH,BV$ and the other one, modes $AV,BH$.

For each of the states $|\Psi_{1,2}\rangle$, the Schmidt number is $K_1=K_2=1+2N_0$, where $N_0\equiv\sinh^2\Gamma$ is the mean photon number. The total Schmidt number is their product.
Note that none of the states $|\Psi_{1,2}\rangle$, taken separately, violates condition (\ref{separability1}), although they both manifest photon-number entanglement. This shows that there is a difference between polarization and photon-number entanglement of macroscopic Bell states.

If there are many ($M$) independent frequency-wavevector mode pairs, each containing a state of the form (\ref{state}), then the total Schmidt number is given by the product, $K=(1+2N_0)^{2M}$, which is extremely large. Under certain experimental conditions this huge amount of entanglement could be used. However, in our experiment we  treat the whole ensemble of modes jointly; moreover, the detection scheme also combines the two wavelengths. With the only partition being the polarization one, the Schmidt decomposition for our state can be written as for a two-mode squeezed vacuum, but with $\lambda_n$ given by the Poissonian distribution~\cite{Silberhorn}  with the mean value $N\equiv 2MN_0$. Then the Schmidt number can be calculated as $K=e^{2N}/I_0(2N)$, where $I_0$ is the zero-order modified Bessel function of the first kind. At large $N, K\approx2\sqrt{\pi N}$. In our experiment, this yields $K\approx 10^3$. This is much less than in the single-mode case because we do not address each mode separately and only deal with their ensemble.
We \color{black} stress that the Schmidt number is not an operational measure as it cannot be directly measured in experiment. However, there exists its operational counterpart.

\textit{\textbf{Operational measure.}} The Schmidt number can be understood as the effective number of the Schmidt modes.
For continuous variables like wavevectors or frequencies of single photons, its operational counterpart was proposed~\cite{R} as the ratio of the unconditional width of one subsystem w.r.t. some parameter to the width of the corresponding conditional distribution.

By analogy, consider the following procedure. In experiment, we obtain the photon-number probability distribution in, say, beam B (\textit{unconditional distribution}). Next, we measure the photon-number \textit{conditional distribution} for beam B by post-selecting only those pulses for which the photon number in beam A is fixed, or within certain narrow bounds. The ratio of the widths for the two distributions can be considered as an operational measure of entanglement. Note that the narrowing of the photon-number conditional distribution is a typical feature of twin beams~\cite{conditional} but has never been considered as a measure of entanglement.

\begin{figure}[h]
\begin{center}
\includegraphics[width=0.3\textwidth]{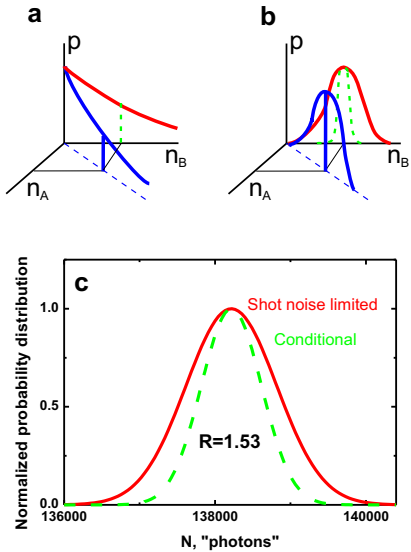}
\caption{Photon-number distributions. (a) single-mode case: the joint probability distribution (blue line in the plane $n_A=n_B$), the unconditional distribution (exponentially decaying red line), and the conditional distribution (green dashed line). (b) the same distributions in the case of a realistic experiment (many modes, non-ideal detection). (c) experimentally obtained distributions: the unconditional one, with the excess noise eliminated  (red solid line) and the conditional one (green dashed line). The electronic noise is subtracted. }\label{Distributions}
\end{center}
\end{figure}

For each pair of correlated modes (for instance, $AH-BV$), the joint probability distribution contains a factor $\delta_{n_A,n_B}$ (Fig.~\ref{Distributions}a). Therefore, while the unconditional distribution for beam $B$ has a negative exponential shape, the conditional probability is only nonzero for a single value of $n_B$. In the multimode case, the unconditional distribution is Poissonian (Fig.~\ref{Distributions}b) while the conditional distribution is again of unity width. This gives $R=2\sqrt{2N\ln2}$, almost coinciding with the Schmidt number. Unfortunately, the conditional distribution gets broadened due to losses, and the resulting $R$ value becomes $R_{\eta}=1/\sqrt{1-\eta}$, where $\eta$ is the overall detection efficiency. Thus, the accessible degree of entanglement is given by only the detection efficiency and turns out to be much less than the Schmidt number.

In our experiment, the unconditional distribution is broadened due to the excess noise of the pump. Therefore we find the numerator of the $R$ ratio by measuring the width of a shot-noise limited source with the same mean photon number (Fig.~\ref{Distributions}c, red solid line).
The degree of entanglement is then measured to be $1.53\pm0.05$. This corresponds to the detection efficiency $\eta=0.57$.


\textit{\textbf{The triplet states.}} We have proved the violation of the
separability condition (\ref{separability1}) for the macroscopic singlet Bell state.
Since the triplet states can be obtained from the singlet one via local unitary transformations, they are entangled as well. The separability conditions for these states can be derived from condition (\ref{separability1}) via the corresponding unitary transformations~\cite{to be}.

The Schmidt decomposition for the triplet states is similar to the one for the singlet state, and the Schmidt number is the same. The only difference is that for the singlet state, the polarization modes can be chosen in any way. For each of the triplet states, there is a unique choice of polarization modes to see entanglement: it should be horizontally and vertically polarized modes for $|\Psi^{(+)}_{mac}\rangle$, diagonally polarized modes for $|\Phi^{(-)}_{mac}\rangle$, and right- and left-circularly polarized modes for $|\Phi^{(+)}_{mac}\rangle$~\cite{Macrobell}.

In conclusion, we have tested the macroscopic singlet Bell state, containing two beams of different wavelengths, for separability. The results convincingly prove that the state is non-separable with respect to polarization observables. As the photon numbers per pulse are as high as $10^5$, and the state is prepared pure, this can be considered as a proof of macroscopic entanglement. As a measure of photon-number entanglement, we have calculated the Schmidt number, which turned out to be given by the total number of photons in a pulse. Theoretically, our multiphoton state is highly entangled. The entanglement can be confirmed by the measurement of the photon-number distribution for one beam (unconditional distribution) and the distribution conditioned on registering a certain number of photons in the other beam (conditional distribution). As the former is broader than the latter, we conclude that the state is photon-number entangled. The measured degree of entanglement and separability condition violation are both reduced due to the inefficient detection. However, there is apparently a difference between photon-number and polarization entanglement.

\textbf{Acknowledgements}
We acknowledge partial financial support of the Russian Foundation for Basic Research, grants 10-02-00202, 11-02-01074, and 12-02-00965.
T.Sh.I. acknowledges funding from Alexander von Humboldt Foundation.
We are grateful to Marek \.Zukowski, Pavel Sekatsky, Magda Stobinska, and Farid Khalili for stimulating discussions and to Ivan Gonoskov for the help in calculations.


\newpage

\begin{center}
    \textbf{Supplementary information: Polarization-Entangled Light Pulses of $10^5$ Photons}
\end{center}

\vspace{5mm}

\subsection{Derivation of the separability condition.}

In this derivation, we will proceed along the same line as Simon and Bouwmeester~\cite{Simon}, with the difference that we will use polarization variables. This makes the proof more consistent with the experimental part. Another important difference is that we go further than the authors of Ref.~\cite{Simon} and prove a stronger inequality, bounding from below not the mean square value of a vector operator but its variance. Finally, we extend the inequality to the multi-mode case.

Consider a bipartite system containing two macroscopic light beams, $A$ and $B$, which are distinguishable in wavelength (our case) or wavevector (the case considered in Ref.~\cite{Simon}). The Stokes vector of such a system, defined as a set of three Stokes operators, $\mathbf{\hat{S}}\equiv\{\hat{S}_1;\hat{S}_2;\hat{S}_3\}$ is given by the sum of the Stokes vectors for beams $A,B$~\cite{Simon,Macrobell},
\begin{equation}
\mathbf{\hat{S}}=\mathbf{\hat{S}}^A+\mathbf{\hat{S}}^B. \label{Stokes_sum}
\end{equation}

Suppose that the beams are separable. Then the density matrix of the total state is $\hat{\rho}=\sum_jp_j\hat{\rho}_j^A\bigotimes\hat{\rho}_j^B$. The square of the Stokes vector operator is $\mathbf{\hat{S}}^2\equiv\hat{S}_1^2+\hat{S}_2^2+\hat{S}_3^2$, and its mean value can be written as
\begin{equation}
\langle\mathbf{\hat{S}}^2\rangle=\sum_jp_j\langle[\mathbf{\hat{S}}^A]^2\rangle_j+\sum_jp_j\langle[\mathbf{\hat{S}}^B]^2\rangle_j+2
\sum_jp_j\langle\mathbf{\hat{S}}^A\rangle_j\langle\mathbf{\hat{S}}^B\rangle_j\nonumber.
\end{equation}
The subscript $j$ by the angular brackets means that the averaging is made over the state $\hat{\rho}_j^{A,B}$. Here, $p_j\ge0$ are classical probabilities, $\sum_jp_j=1$. It is important that in the last term, the averaging goes separately for each subsystem due to separability, $\langle\mathbf{\hat{S}}^A\mathbf{\hat{S}}^B\rangle_j=
\langle\mathbf{\hat{S}}^A\rangle_j\langle\mathbf{\hat{S}}^B\rangle_j$.

The variance of $\mathbf{\hat{S}}$ takes the form
\begin{eqnarray}
\mathbf{\Delta\hat{S}}^2\equiv\langle\mathbf{\hat{S}}^2\rangle-\langle(\mathbf{\hat{S}}^A+\mathbf{\hat{S}}^B)\rangle^2=\nonumber\\
\sum_jp_j\langle[\mathbf{\hat{S}}^A]^2\rangle_j+\sum_jp_j\langle[\mathbf{\hat{S}}^B]^2\rangle_j+2
\sum_jp_j\langle\mathbf{\hat{S}}^A\rangle_j\langle\mathbf{\hat{S}}^B\rangle_j\nonumber\\
-\left(\sum_jp_j\langle\hat{S}_1^A+\hat{S}_1^B\rangle_j\right)^2-
\left(\sum_jp_j\langle\hat{S}_2^A+\hat{S}_2^B\rangle_j\right)^2\nonumber\\
-\left(\sum_jp_j\langle\hat{S}_3^A+\hat{S}_3^B\rangle_j\right)^2\nonumber.
\end{eqnarray}
The last three terms here are squared mean values calculated by averaging with usual classical probabilities $p_j\ge0$. This implies the positivity of the variance, and for each term we can write
\begin{equation}
\left(\sum_jp_j\langle\hat{S}_k^A+\hat{S}_k^B\rangle_j\right)^2\le
\sum_jp_j\langle\hat{S}_k^A+\hat{S}_k^B\rangle_j^2,\,\,k=1,2,3\nonumber,
\end{equation}
which leads us to an inequality for the variance,
\begin{eqnarray}
\mathbf{\Delta\hat{S}}^2\ge
\sum_jp_j\langle[\mathbf{\hat{S}}^A]^2\rangle_j+\sum_jp_j\langle[\mathbf{\hat{S}}^B]^2\rangle_j+2
\sum_jp_j\langle\mathbf{\hat{S}}^A\rangle_j\langle\mathbf{\hat{S}}^B\rangle_j\nonumber\\
-\sum_jp_j\langle\hat{S}_1^A+\hat{S}_1^B\rangle_j^2-
\sum_jp_j\langle\hat{S}_2^A+\hat{S}_2^B\rangle_j^2
-\sum_jp_j\langle\hat{S}_3^A+\hat{S}_3^B\rangle_j^2\nonumber.
\end{eqnarray}

Again, due to separability, $\langle\hat{S}_1^A\hat{S}_1^B\rangle_j+\langle\hat{S}_2^A\hat{S}_2^B\rangle_j+\langle\hat{S}_3^A\hat{S}_3^B\rangle_j\equiv
\langle\mathbf{\hat{S}}^A\mathbf{\hat{S}}^B\rangle_j\equiv\langle\mathbf{\hat{S}}^A\rangle_j\langle\mathbf{\hat{S}}^B\rangle_j$. This leads to the inequality
\begin{eqnarray}
\mathbf{\Delta\hat{S}}^2\ge\sum_jp_j\langle[\mathbf{\hat{S}}^A]^2\rangle_j+
\sum_jp_j\langle[\mathbf{\hat{S}}^B]^2\rangle_j\nonumber\\
-\sum_jp_j\langle\mathbf{\hat{S}}^A\rangle_j^2-\sum_jp_j\langle\mathbf{\hat{S}}^B\rangle_j^2\nonumber.
\end{eqnarray}
To each beam $A,B$ taken separately, we can apply the well-known operator equality~\cite{Klyshko},
\begin{equation}
[\mathbf{\hat{S}}^
{A,B}]^2=\hat{S}_{0}^{A,B}(\hat{S}_{0}^{A,B}+2), \label{sep}
\end{equation}
which yields
\begin{eqnarray}
\mathbf{\Delta\hat{S}}^2\ge
\sum_jp_j\langle\hat{S}_0^A(\hat{S}_0^A+2)\rangle_j+\sum_jp_j\langle\hat{S}_0^B(\hat{S}_0^B+2)\rangle_j\nonumber\\
-\sum_jp_j\langle\mathbf{\hat{S}}^A\rangle_j^2-\sum_jp_j\langle\mathbf{\hat{S}}^B\rangle_j^2\nonumber.
\end{eqnarray}

Note that up to this step, no assumption was made that the beams $A,B$ are single-mode. However, equality (\ref{sep}) is only valid for a single-mode beam. Hence, so far we restrict our consideration to a single pair of modes. The multi-mode case will be considered below.

 Since $\langle\mathbf{\hat{S}}^{A,B}\rangle_j^2\equiv P_j^{A,B}\langle\hat{S}_{0}^{A,B}\rangle_j^2$, where $0\le P_j^{A,B}\le1$ is the first-order degree of polarization~\cite{Shurcliff}, the inequality can be rewritten as
\begin{eqnarray}
\mathbf{\Delta\hat{S}}^2\ge
\sum_jp_j\langle\hat{S}_0^A(\hat{S}_0^A+2)\rangle_j+\sum_jp_j\langle\hat{S}_0^B(\hat{S}_0^B+2)\rangle_j\nonumber\\
-\sum_jp_j\langle\hat{S}_0^A\rangle_j^2-\sum_jp_j\langle\hat{S}_0^B\rangle_j^2\nonumber.
\end{eqnarray}
Note that $\langle[\hat{S}_{0j}^A]^2\rangle\ge\langle\hat{S}_{0j}^A\rangle^2$ (the variance is non-negative), and $\sum_jp_j\langle\hat{S}_0^A+\hat{S}_0^B\rangle_j=\langle\hat{S}_0\rangle$, hence
we get the separability condition in the form
\begin{equation}
\Delta\mathbf{\hat{S}}^2\ge2\langle\hat{S}_0\rangle. \label{separability}
\end{equation}
A similar inequality was proved for separable atomic ensembles in Ref.~\cite{Toth}. This inequality is stronger than the separability condition derived in Ref.~\cite{Simon},
\begin{equation}
\langle\mathbf{\hat{S}}^2\rangle\ge2\langle\hat{S}_0\rangle.\label{sepSimon}
 \end{equation}
 In the special case of light unpolarized in the first order, for which $\langle\mathbf{\hat{S}}\rangle=0$, both conditions coincide.

It should be emphasized that inequality (\ref{separability}), as well as inequality (\ref{sepSimon}), follows from the separability assumption and is therefore \emph{a necessary condition of separability}. Hence, the opposite inequalities are \emph{sufficient conditions of non-separability} and can be called \emph{non-separability witnesses}.

In the multimode case, where each beam $A,B$ contains a set of independent modes, one can write inequality (\ref{separability}) for each pair of modes $A_k,B_k$,
\begin{equation}
\Delta\mathbf{\hat{S}}_k^2\ge2\langle\hat{S}_0\rangle_k. \label{sepmode}
\end{equation}
The variance of any observable for an ensemble of independent modes is equal to the sum of separate variances, $\Delta\mathbf{\hat{S}}^2=\sum_k\Delta\mathbf{\hat{S}}_k^2$, and the same is true for the mean values, $\langle\hat{S}_0\rangle=\sum_k\langle\hat{S}_0\rangle_k$. Then, by summing inequalities (\ref{sepmode}) we again obtain inequality (\ref{separability}) but this time it is valid for a multi-mode system. Therefore, the opposite inequality,
\begin{equation}
\Delta\mathbf{\hat{S}}^2<2\langle\hat{S}_0\rangle, \label{witness}
\end{equation}
is a witness of non-separability for any kind of states, including mixed and multi-mode ones.

\subsection{Preparation of the $|\Psi^{(-)}_{mac}\rangle$ state}
\begin{figure}[h]
\includegraphics[width=0.35\textwidth]{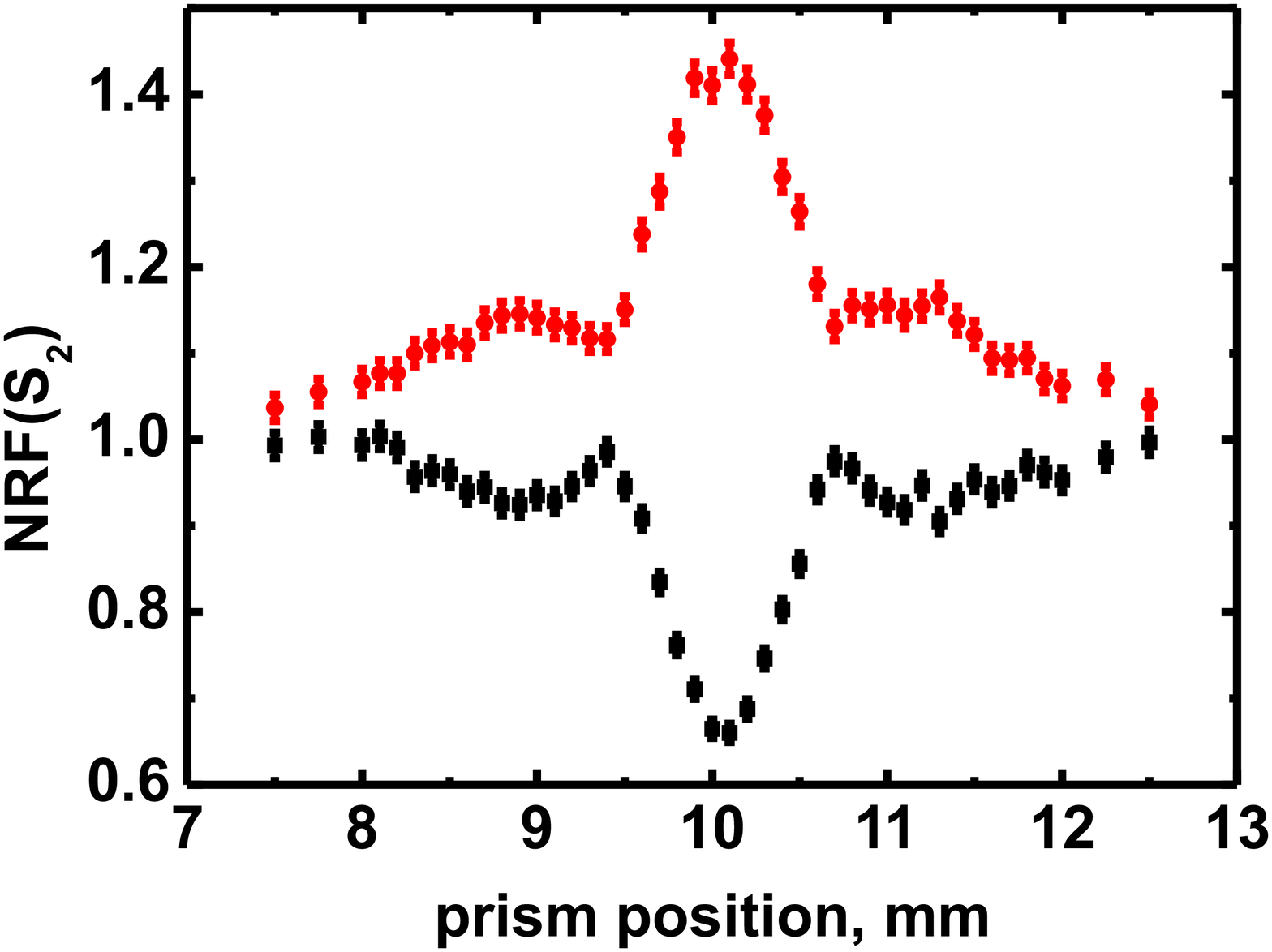} \caption{Interference pattern envelope (minimum and maximum NRF values
for the $S_2$ Stokes observable) versus the path length difference between the arms of the MZ interferometer.}
\label{fig1s}
\end{figure}

We pump two orthogonally oriented type-I BBO crystals placed in two arms of a Mach-Zehnder interferometer (MZI) with a strong pulsed beam (third harmonic of Nd:YAG laser with the pulse duration $18$ ps, pulse energy up to $0.2$ mJ, and repetition rate $1$ kHz). The four-mode state is realized by, first, using frequency non-degenerate phase matching, signal and idler wavelengths being $\lambda_A=635$ nm and $\lambda_B=805$ nm, respectively, and, second, overlapping two orthogonally polarized squeezed vacuum beams on a polarizing beamsplitter. The phase between the beams is set to be $\pi$ using a piezoelectric feed; it is important that the MZI is balanced with an accuracy better than the coherence time of the pump, which is $5$ ps. To provide this, we put one of the interferometer mirrors on a
piezoelectric feed combined with a micrometric translation stage. Fig.~\ref{fig1s} shows how the interference
visibility depends on the `rough' shift of the micrometric stage.
The interference is observed by measuring the $S_2$ Stokes variance versus the bias voltage applied
to the piezoelectric feed. The bias voltage changes the phase $\varphi$ between the squeezed vacuums produced in
the MZ interferometer, so that the state at the output is
\begin{equation}\label{state}
|\Psi(\varphi)\rangle=\exp[\Gamma(a_{H}^{\dagger} b_{H}^{\dagger}+e^{i\varphi}
a_{V}^{\dagger}b_{V}^{\dagger})+\hbox{h.c.}]|\hbox{vac}\rangle.
\end{equation}
\begin{figure}[h]
\includegraphics[width=0.4\textwidth]{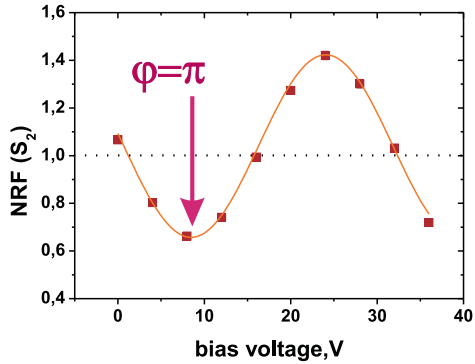} \caption{NRF of the Stokes observable $S_2$ as a function of
the bias voltage at the piezoelectric feed. The arrow shows the point where  the fluctuations of $S_2$ are
suppressed, which indicates the production of $|\Phi^{(-)}_{mac}\rangle$ and hence, the phase $\varphi=\pi$.}
\label{fig2s}
\end{figure}
At phase $\varphi=\pi$, the state us $|\Phi^{(-)}_{mac}\rangle$, and it has reduced fluctuations of the $S_2$
observable~\cite{Macrobell}. At $\varphi=0$, the state is $|\Phi^{(+)}_{mac}\rangle$ and the fluctuations of $S_2$ are enhanced. Scanning of the bias voltage and hence the phase $\varphi$ creates a sine
variation of the normalized variance (noise reduction factor, NRF) of $S_2$ (Fig.~\ref{fig2s}). By plotting the
minimal and maximal NRF values versus the reading of the micrometric stage we get the envelope of the
interference pattern (Fig.~\ref{fig1s}).

The residual pump is eliminated using a dichroic mirror in each arm and a colour-glass filter after the MZI. The output state $|\Phi^{(+)}_{mac}\rangle$ becomes $|\Psi^{(+)}_{mac}\rangle$ in the $45^{\circ}$-tilted basis. It is turned into the singlet state $|\Psi^{(-)}_{mac}\rangle$ with the help of a dichroic waveplate (see the next section).

It is important that, for better mode matching, beams of different wavelengths should be filtered in the transverse wavevector by different apertures, with the diameters proportional to the wavelengths. We satisfy this condition by separating the wavelengths at the output of the MZI, with the help of a dichroic mirror, and using a separate iris aperture for each wavelength. Both apertures are placed in the focal plane of a lens with the focal length $300$ mm. The crystals, in their turn, are placed at a distance of $300$ mm from the lens. This way, the aperture diameters can be adjusted independently so that the same range of transverse wavevector (the same numbers of transverse modes) for both signal and idler wavelengths can be filtered out. The $635$ nm and $805$ nm beams are then joined on another dichroic mirror, and for the whole beam the Stokes measurement is performed. Note that when two beams of different colors are registered by the same detectors, the Stokes observables simply add up, as the detectors do not resolve the frequency beats.

Our detectors are p-i-n diodes followed by charge-sensitive amplifiers. For each light pulse, they both produce output pulses with a fixed shape and the amplitude scaling linearly with the total number of registered electrons. The output pulses are time-integrated by means of an AD card. The electronic noise is mainly caused by the electric circuit as well as the digitization noise of the AD card and amounts to $300$ electrons/pulse. Before the measurement, the detectors are calibrated using a shot-noise limited source (strongly attenuated laser radiation), by measuring the variance of the difference signal $S_-$ as a function of the mean sum signal $\langle S_+\rangle$. In the range of $\langle S_+\rangle$ on the order of $10^4$ - $10^6$ photons per pulse, the dependence is linear, $\Delta S_-^2=\alpha\langle S_+\rangle+\Delta S_e^2$, $\Delta S_e^2$ being the electronic noise contribution. The first term of this dependence is considered as the shot-noise level and used in the calculation of NRF.

\subsection{Transformation performed by the dichroic plate.}
The transformation performed by the dichroic plate can be described as phase shifts introduced into the creation and annihilation operators, different for the two wavelengths and two polarizations:
\begin{eqnarray}
a_H\rightarrow a'_H=a_H e^{i\phi_{AH}}, \,\,a_V\rightarrow a'_V=a_V e^{i\phi_{AV}},\nonumber\\
b_H\rightarrow b'_H=b_H e^{i\phi_{BH}}, \,\,b_V\rightarrow b'_V=b_V e^{i\phi_{BV}},\nonumber\\
\label{shifts}
\end{eqnarray}
where
\begin{equation}
\phi_{AH}\equiv k_A^od,\,\phi_{AV}\equiv k_A^e d, \label{phases}
\end{equation}
and similarly for the $B$ beam. Here, $k_{A,B}^{o,e}$ are ordinary (extraordinary) wavevectors for beams $A,B$ and $d$ is the plate thickness.

The dichroic plate is a $170\mu$ crystal quartz plate. The ordinary and extraordinary wavevectors $k_{A,B}^{o,e}$ for quartz can be calculated using the Sellmeier equations (see, for instance,~\cite{Sellmeier}). As a result, we get the o-e phase delay for $\lambda_A$ equal to $\phi_{AH}-\phi_{AV}=4.854\pi$ and for $\lambda_B$, $\phi_{BH}-\phi_{BV}=3.774\pi$, their difference being $1.08\pi\approx\pi$.

Then the transformation (\ref{shifts}) turns the Hamiltonian generating the triplet state $|\Psi^{(+)}_{mac}\rangle$,
\begin{equation}
\hat{H'}=i\hbar G(a_{H}^{\dagger}b_{V}^{\dagger}+a_{V}^{\dagger}b_{H}^{\dagger})+\hbox{h.c.},
 \label{Ham+}
\end{equation} into
\begin{eqnarray}
\hat{H''}=i\hbar G e^{i(\phi_{AH}+\phi_{BV})}\nonumber\\
\times(a_{H}^{\dagger}b_{V}^{\dagger}+e^{i(\phi_{BH}-\phi_{BV}-\phi_{AH}+\phi_{AV})}
a_{V}^{\dagger}b_{H}^{\dagger})\nonumber\\
+\hbox{h.c.}\approx i\hbar G e^{i(\phi_{AH}+\phi_{BV})}(a_{H}^{\dagger}b_{V}^{\dagger}-a_{V}^{\dagger}b_{H}^{\dagger})+\hbox{h.c.},
 \label{Ham-}
\end{eqnarray}
which, up to a phase factor, coincides with the Hamiltonian $\hat{H}$ generating the singlet state $|\Psi^{(-)}_{mac}\rangle$.


\begin{references}

\bibitem{DNK98}
D.~N.~Klyshko, Physics - Uspekhi \textbf{41}, (1998) 885-922.


\bibitem{BEC} J. Este`ve et al.,
Nature \textbf{455}, 1216-1219 (2008).

\bibitem{Polzik}B.~Julsgaard, A.~Kozhekin, and E.~S.~Polzik, Nature \textbf{413}, 400 (2001).

\bibitem{macroqubits} F.~De Martini, F.~Sciarrino, and C.~Vitelli, PRL \textbf{100}, 253601 (2008); F.~De Martini et al., PRL \textbf{104}, 050403 (2010); N. Spagnolo et al., PRA \textbf{85}, 052101 (2010).

\bibitem{Gisin}P. Sekatski et al.,
PRL \textbf{103}, 113601 (2009).

\bibitem{Drummond}P.~D.~Drummond, PRL \textbf{50}, 407 (1983).

\bibitem{Reid}W.~J.~Munro and M.~D.~Reid, PRA \textbf{47}, 4412 (1993).

\bibitem{Gatti} A.~Gatti et al., PRA \textbf{68}, 053807 (2003).

\bibitem{Durkin} G.~A.~Durkin et al., PRA \textbf{70}, 062305 (2004).

\bibitem{Simon}Ch.~Simon and D.~Bouwmeester, PRL~\textbf{91}, 053601 (2003).

\bibitem{Eisenberg}H.~S.~Eisenberg et al., PRL~\textbf{93}, 193901 (2004).

\bibitem{Lamas-Linares} A.~Lamas-Linares, J.~C.~Howell, and D.~Bouwmeester, Nature~\textbf{412}, 887 (2001).

\bibitem{Macrobell}T.~Sh.~Iskhakov et al.,
PRL \textbf{106}, 113602 (2011).

\bibitem{Braunstein} S.~L.~Braunstein, PRA \textbf{71}, 055801 (2005).

\bibitem{qmemory} L.~V.~Gerasimov et al.,
arXiv:1111.6669v1 [quant-ph] 29 Nov 2011.

\bibitem{diTrapani}O.~Jedrkiewicz et al.,
 Optics Express \textbf{19}, 12903 (2011).

\bibitem{supplementary} See the supplementary information.

\bibitem{Toth} G.~T\'oth, Phys. Rev. A.~\textbf{69}, 052327 (2004).

\bibitem{Klyshko} D.~N.~Klyshko, JETP \textbf{84}, 1065 (1997).

\bibitem{Klimov} A.~B. Klimov et al.,
PRA \textbf{71}, 033813 (2005).


\bibitem{Karas}V.~P.~Karassiov, J. Phys. A \textbf{26}, 4345 (1993).

\bibitem{Karas2}V.~P.~Karassiov, J. of Russian Laser Research,  \textbf{26}, 484 (2005).

\bibitem{Iskhakov} T.~Sh.~Iskhakov, M.~V.~Chekhova, and G.~Leuchs, PRL~\textbf{102}, 183602 (2009).

\bibitem{Fedorov} M.~V.~Fedorov et al., PRL~\textbf{99}, 163901 (2007).

\bibitem{two-color}I.~N.~Agafonov, M.~V.~Chekhova, and G.~Leuchs, PRA \textbf{82}, 011801 (2010).

\bibitem{R} M.~V.~Fedorov et al., PRA \textbf{69}, 052117 (2004).

\bibitem{Gisin2} N. Gisin, Phys. Lett. A \textbf{154}, 201 (1991); N. Gisin and A. Peres, Phys. Lett. A \textbf{162}, 15-17 (1992); S. Popescu and D. Rohrlich, Phys. Lett. A \textbf{166}, 293 (1992).

\bibitem{conditional} S.~Sp\"alter et al., PRL \textbf{81}, 786 (1998); M.~Bondani et al., Phys. Rev. A \textbf{76}, 013833 (2007).

\bibitem{to be} Pavel Sekatsky, private communication.

\bibitem{Silberhorn} M.~Avenhaus et al., PRL \textbf{104}, 063602 (2010).


\end{references}

\begin{references}

\bibitem{Simon}Ch.~Simon and D.~Bouwmeester, Phys. Rev. Lett.~\textbf{91}, 053601 (2003).

\bibitem{Macrobell}T.~Sh.~Iskhakov et al.,
PRL \textbf{106}, 113602 (2011).

\bibitem{Klyshko} D.~N.~Klyshko, JETP \textbf{84}, 1065 (1997).

\bibitem{Shurcliff} W.~A. Shurcliff, \textit{Polarized Light: Production and Use}, Harvard Univ. Press, Cambridge, MA (1962).

\bibitem{Toth} G.~T\'oth, Phys. Rev. A.~\textbf{69}, 052327 (2004).

\bibitem{two-color} I.~N.~Agafonov, M.~V.~Chekhova, and G.~Leuchs, Phys. Rev. A.~\textbf{82}, 011801(R) (2010).

\bibitem{Sellmeier} http://www.cvimellesgriot.com/products/Documents/Catalog/ Dispersion\_Equations.pdf

\end{references}
\end{document}